\documentclass[12pt,a4paper]{article}
\usepackage{amsthm}
\theoremstyle{plain}
\usepackage{authblk}
\usepackage[english]{babel}
\usepackage{graphicx}
\usepackage[latin1]{inputenc}
\usepackage{verbatim}
\usepackage{amsfonts}
\usepackage{amsmath}
\usepackage{txfonts}
\usepackage[T1]{fontenc}
\usepackage{color}
\usepackage{epsfig}
\usepackage{natbib}
\usepackage{tikz-cd}
\usepackage{url}
\usepackage[bookmarksnumbered=true, colorlinks=true, citecolor=blue]{hyperref}
\date{}

\title{Does General Relativity Highlight Necessary Connections in Nature?}
\author{Antonio Vassallo}
\affil{Warsaw University of Technology\\Faculty of Administration and Social Sciences\\Plac Politechniki 1\\ 00-661 Warsaw\\ \url{antonio.vassallo1977@gmail.com}}

\begin{document}

\maketitle

\begin{center}
Accepted for publication in \emph{Synthese}
\end{center}

\pdfbookmark[1]{Abstract}{abstract}
\begin{abstract}
The dynamics of general relativity is encoded in a set of ten differential equations, the so-called \emph{Einstein field equations}. It is usually believed that Einstein's equations represent a physical law describing the coupling of spacetime with material fields. However, just six of these equations actually describe the coupling mechanism: the remaining four represent a set of differential relations known as \emph{Bianchi identities}. The paper discusses the physical role that the Bianchi identities play in general relativity, and investigates whether these identities --\emph{qua} part of a physical law-- highlight some kind of \emph{a posteriori} necessity in a Kripkean sense. The inquiry shows that general relativistic physics has an interesting bearing on the debate about the metaphysics of the laws of nature.\\

\end{abstract}

\section{Introduction}\label{sec:1}
The debate on the metaphysics of the laws of nature is huge. Philosophers do not agree on whether there exist such things as laws of nature (see, e.g., \citealp{584}, for a radical antirealist position), let alone on what they are supposed to be. In the present paper, I am going to circumscribe this immense conceptual space by making two main working hypotheses. The first is that, in fact, there exist such things as laws of nature, leaving open whether they are fundamental or reducible to more fundamental features of reality. The second is that laws of nature properly said are \emph{scientific} laws and, more specifically, the laws of fundamental physics --basically, the physicalists' dream. One may or may not be sympathetic with these two claims but, I hope, no one will contest the fact that they are at least plausible assumptions.

However, this is still not enough to enclose the conceptual space into a tractable area. In fact, we might encounter huge complications even by restricting ourselves to the laws of fundamental physics. This is because  the laws of quantum theory --one of the pillars of modern physics-- enjoy a rather unclear metaphysical status (to put it mildly). Indeed, some claim that quantum theory \emph{as it stands} is a genuine description of how things behave in the world, while others are convinced that it is not even a full-fledged theory, but just an algorithm for extracting experimental predictions (see \citealp{585}, for an introduction to this debate and a defense of the second position). Since I do not want this paper to get roped in the debate about the interpretations of quantum theory, I will instead focus on the other pillar of modern physics, namely, general relativity.

A final disclaimer: Throughout the text I am going to use the possible-world terminology just as a useful conceptual tool, without committing myself to any particular metaphysical stance regarding possible worlds.

Now that the field of inquiry has been reasonably restricted, I am ready to ask the research question this paper is concerned with: Are the laws of general relativity \emph{necessary} in any non-trivial, interesting sense? The way I am formulating the question automatically cuts off from the picture instances of \emph{logical} or \emph{conceptual} necessity (e.g. a bachelor is necessarily, that is, by definition, an unmarried man) and \emph{nomic} necessity (e.g. a physical body inhabiting a world where the physical law $L$ holds, necessarily behaves in accordance with $L$). What I am most interested in is, instead, whether Einstein's equations bring to the fore any instance of what Kripke would call \emph{a posteriori} necessity, that is, necessary truths that we can discover only by empirical investigation (cf. \citealp{288}, especially pages 140-144). 

Kripke famously argued that the truth of a certain class of statements, including identity statements (e.g. ``Water is H$_2$O'', ``Phosphorus is Hesperus'') and natural kinds characterization (e.g. ``Potassium is an alkali metal'', ``Cats are mammals''), is necessary in a sense close to that of statements such as ``Bachelors are unmarried men'', yet not \emph{a priori} in that such a truth has to be ascertained by looking at the outside world. This type of necessity is enthralling to metaphysicians since it highlights some sort of constraint on reality itself --i.e. some things being ``defined'' to be as they are and not otherwise.

Proponents of the regularity view of laws, in particular the Humeans, forcefully deny that laws of nature possess any necessary connotation tout-court, let alone the peculiar type envisaged by Kripke. For them, laws supervene on the contingent arrangement of (local) states of affairs making up a world (see \citealp{510}, for a self-contained introduction to this stance). Hence, they deem entirely possible to have a world where the laws of physics allow for potassium being in fact a nonmetal.\footnote{As philosophically controversial as this view might be, it certainly makes for terrific science fiction: \citet{590} is a wondrous example.}

Surprisingly enough, this skepticism towards the necessity of laws of nature is shared, though with some substantial divergencies, by ``universals theorists'' \emph{\`a la} Armstrong (e.g. \citealp{583}). Contrary to regularity theorists, universals theorists do claim that the fact that it is a law in a world $w$ that, say, all $F$s are $G$s  means that some sort of ``necessitation'' relation $N$ holds between the universals ``$F$-ness'' and ``$G$-ness''. However, such a necessity just boils down to the fact that \emph{if} $F$ is instantiated at $w$ \emph{and} $N(F,\cdot)$ holds, \emph{then} the second \emph{relatum} of $N$ is necessarily $G$. Obviously, this does not imply that ``$N(F,G)$'' has to hold in all possible worlds, not least because otherwise the universals of ``$F$-ness'' and ``$G$-ness'' would become necessary beings (see \citealp{583}, chapter 11, and in particular section 2, for a justification of this claim). Hence, although universals theorists allow for some sort of necessity being at play in the laws of nature, they nonetheless conceive of this necessitation as being of the nomic kind or a variant thereof (and, for sure, much weaker than the one envisaged by Kripke).

However, there is an important reason to be suspicious of both regularity an universals theories of laws, this reason being that both theories have troubles to show how laws of nature support counterfactual reasoning (\citealp{591}, neatly summarizes such concerns). In the regularity theory case, the fact that all $F$s are $G$s is a law means that it is contingently true that ``$\forall x\left(\mathcal{F}(x)\Rightarrow \mathcal{G}(x)\right)$''. If we denote by $Ext(\mathcal{P})$ the extension of the predicate corresponding to the property $P$, then the previous law statement just means that $Ext(\mathcal{F})$ is included in $Ext(\mathcal{G})$. Does this law support a counterfactual of the form ``if an individual $a$ had been $F$, then it would have been $G$''? We immediately see the trouble here: if we enlarge the content of $Ext(\mathcal{F})$ to include the counterfactual case where $a$ is $F$, we are substantially changing the facts upon which the truth of the above law statement --\emph{qua} contingent generalization-- rests.

Universals theorists usually defuse the above challenge by pointing out that, even if we add the ``first-order'' fact that $a$ is $F$, we are not altering in any way the ``second-order'' fact that $F$ and $G$ are $N$-related, against which the counterfactual has to be evaluated. But there is another trouble that suggests itself at this point. If we take ``$N(F,G)$'' to be only contingently true, then the way this law supports counterfactuals of the form ``if something had been $F$, then it would have been $G$'' is rather feeble: Indeed, \emph{if} $N(F,G)$ holds, \emph{then} the counterfactual is true. Otherwise said, in order to evaluate this statement we have to hold fixed not only all the particular aspects relevant to the situation in question, but also the law itself! This makes the evaluation procedure suspiciously look like a matter of convention rather than the assessment of an objective modal fact.

These troubles with counterfactuals are one of the reasons why some philosophers entertain the much stronger idea that laws of nature are metaphysically necessary in a Kripkean sense. For these people, once (if?) we will get to know the true laws that govern our world, we will realize that they could not have been otherwise. In other words, the laws of nature for these strong necessitarians constrain not only physical but also metaphysical possibilities. This idea is generally implemented through a causal theory of properties. For example, \citet{588}, puts it this way:

\begin{quote}
[T]he claim of the causal theory of properties is that the properties that have causal features \emph{non}-derivatively have them essentially, and are individuated in terms of them. [footnote suppressed] I think this comes to much the same thing as saying that the properties that enter into causal laws have their causal features essentially, and are individuated in terms of them.\\
(\emph{ibid.}, page 65)
\end{quote}

\begin{quote}
So insofar as causal laws can be construed as describing the causal features of properties, they are necessary truths. One way to get the conclusion that laws are necessary is to [adopt the view] that laws are, or assert, relations between properties.\\
(\emph{ibid.}, page 61)
\end{quote}

Under this view, the claim that potassium is an alkali metal highlights a necessary connection in nature in the sense that it is essential to potassium to behave as an alkali metal and thus, for example, to violently interact with oxygen under certain circumstances $X$. Note how this causal interaction between potassium and oxygen can be rendered in terms of a relation holding between (an appropriate subset of) properties borne by these two elements. A counterfactual situation which holds fixed the conditions $X$ but in which potassium does not interact with oxygen is not only physically but also metaphysically impossible. Note also how this counterfactual does not need the laws to be fixed as a matter of convention: Given that they hold in all possible worlds, they are fixed as an objective modal fact.

The details through which this strong necessitarian strategy is implemented may vary from author to author. For example, Shoemaker (\citealp{592}) maintains that the categorical basis of a property is distinct from its intrinsic causal power, while Bird (cf. for example, \citealp{586}) endorses dispositional monism. Ellis, instead, develops an essentialist account of the properties of natural kinds \citep{149}. Moreover, some authors --such as \citet{587}-- argue in favor of a conceptual link between strong necessitarianism and Platonism.

For simplicity's sake, here I will lump together all these variants under a unique strong necessitarian view on laws of nature, and rephrase my original research question as: To what extent --if any-- do the laws of general relativity fulfill the wishes of strong necessitarians?

\section{The inciting incident: The Bianchi identities}\label{sec:2}
The structure of the Einstein field equations (written in natural units such that $G=c=1$) in an arbitrary coordinate system $\left\{x^\mu\right\}$ is:\footnote{Here, for simplicity's sake, I am disregarding the term featuring the cosmological constant.}
\begin{equation}\label{efe}
R_{\mu\nu}-\frac{1}{2}g_{\mu\nu}R=8\pi T_{\mu\nu}[\Phi, g_{\mu\nu}]\quad (\mu,\nu=0,1,2,3).
\end{equation}

The term on the right-hand side features the stress-energy tensor $T_{\mu\nu}$, which depends on the metric tensor $g_{\mu\nu}$ and encodes information about the material distribution of the field(s) symbolized by $\Phi$. For example, the mass-energy density of $\Phi$ as measured by an arbitrary observer with $4$-velocity $u^{\mu}$ is $\rho_{\Phi}=u^{\mu}T_{\mu\nu}u^{\nu}$.

The stress-energy tensor satisfies the following important requirement:

\begin{equation}\label{cons}
T^{\mu\nu}_{\phantom{{\mu\nu}};\mu}=0,
\end{equation}

where the semicolon indicates covariant differentiation with respect to the subsequent index. The expression \eqref{cons} can be seen as a weak local conservation law for the energy-momentum of the material distribution. Roughly speaking, \eqref{cons} states that the amount of $\Phi$'s energy-momentum enclosed in any infinitesimal volume of spacetime is conserved. I called it \emph{weak} conservation law because, less roughly speaking, it is more of a ``balance'' law. To see this, we can rewrite \eqref{cons} \emph{in extenso} using the definition of covariant derivative (a comma indicates ordinary differentiation):

\begin{equation}\label{cons2}
T^{\mu\nu}_{\phantom{\mu\nu};\mu}=T^{\mu\nu}_{\phantom{\mu\nu},\mu}+\Gamma^{\mu}_{\phantom{\mu}\rho\mu}T^{\rho\nu}+\Gamma^{\nu}_{\phantom{\nu}\rho\mu}T^{\mu\rho}= 0,
\end{equation}

where the second and third term of the sum depends on the components of the connection (called \emph{Christoffel symbols}), which defines the covariant derivative. In short, this relation just gives us a measure of how much the covariant derivative of the stress-energy tensor differs from the ordinary one. It is interesting to point out that, in a flat background spacetime, \eqref{cons2} can be interpreted as a measure of how much energy-momentum \emph{fails} to be conserved due to the presence of a gravitational field. However, in general relativity, the ``imbalance'' due to the extra terms in the sum is geometrized away as the curvature of spacetime (which means that the ``real'' divergence of $T_{\mu\nu}$ is given by the covariant derivative), so that the intuitive picture of energy-momentum being conserved in any infinitesimal volume of (curved) spacetime holds (but see \citealp{558}, especially section 3, for a skeptical take on this ``received view''). Note however that, because of this discrepancy between the covariant and the ordinary divergence of the stress-energy tensor, the conservation law \eqref{cons2} cannot be converted into an integral law by using Gauss theorem --which holds for ordinary differentiation-- and, hence, it cannot be extended to finite spacetime regions, let alone to spacetime as a whole (unless spacetime exhibits a highly symmetric structure).

Coming back to Einstein's equations, the term on the left-hand side of \eqref{efe} encodes information about the geometry of spacetime. It is a linear combination of different contractions of the Riemann curvature tensor $R_{\mu\nu\sigma\rho}$, which by itself depends on the metric tensor $g_{\mu\nu}$. This expression on the left-hand side also defines the Einstein tensor $G_{\mu\nu}$:

\begin{equation}\label{eins}
G_{\mu\nu}=R_{\mu\nu}-\frac{1}{2}g_{\mu\nu}R.
\end{equation}

In short, Einstein's equations describe how a material distribution over a region of spacetime (or even the whole of spacetime, in the context of cosmology) influences the geometry of spacetime (more precisely, its curvature) over that region and, in turn, how such a geometry constrains material motions in this region through the geodesic equations of motion entailed by \eqref{efe} (cf., for example., \citealp{27}, section 20.6, for a formal derivation of the equations of motion for test particles from the field equations).

It is easy to realize that \eqref{efe} is a rather dense way to summarize the laws of general relativity. Indeed, given that the tensors appearing there can be written as $4\times4$ matrices, \eqref{efe} can be ``unpacked'' into $16$ equations. However, the matrices $|G_{\mu\nu}|$ and $|T_{\mu\nu}|$ are symmetric, that is, for any element $a_{\mu\nu}$ of them (each element being a function of spacetime points), it is the case that $a_{\mu\nu}=a_{\nu\mu}$. This means that just $10$ Einstein's equations are really needed, the other $6$ being just redundant.

Another moment of reflection shows, however, that this can't be right. If we really had ten equations in ten unknowns $g_{\mu\nu}$, that would mean that --just to have a rough idea-- we could set up an initial value formulation of the dynamics encoded in \eqref{efe} whose initial data would \emph{uniquely fix} the components of the metric tensor   throughout the dynamical evolution --and, hence, uniquely fix a privileged coordinate system for the dynamical description-- (see \citealp{100}, section 10.2, to catch a glimpse of the high non-triviality of the initial value problem in general relativity). This would be a blatant violation of the general covariance of the theory, that is, the fact that any solution of \eqref{efe} can be specified up to an arbitrary choice of four functions that represent a coordinate transformation $x^\mu\longrightarrow x^\nu$ (which is a fancy way to say that we are free to write and solve \eqref{efe} in any coordinate system without loss of physical information).\footnote{The expert readers already know that there is much more to say about the topic of general covariance in general relativity. In section \ref{sec:new}, I will lift just a bit the lid of this Pandora's box.} Fortunately, the issue dissolves once we inspect the $10$ equations and reveal that just $6$ of them relate the spacetime geometry with the material distribution. The other $4$ are mathematical relations involving the Riemann curvature tensor, which go under the name of \emph{Bianchi identities}:

\begin{equation}\label{bian}
R_{\mu\nu\sigma\rho;\delta}+R_{\mu\nu\delta\sigma;\rho}+R_{\mu\nu\delta\rho;\sigma}=0;
\end{equation}
this ``liberates'' the four functions $x^\nu(x^\mu)$ that can hence be freely specified in the initial data without altering the physical information encoded in a solution of \eqref{efe}. This is the point where the physicist nods while the metaphysician raises the eyebrow.

At first sight, it seemed that \eqref{efe} straightforwardly described the coupling of spacetime geometry with matter. Given that many different combinations $\langle g_{\mu\nu},T_{\mu\nu}\rangle$ --again, up to a coordinate transformation-- are compatible with \eqref{efe}, this might have led us to the conclusion that Einstein's equations depict the \emph{contingent} coupling mechanism between geometry and matter. Otherwise said, each (equivalence class of solutions under coordinate transformations) $\langle g_{\mu\nu},T_{\mu\nu}\rangle$ can be taken to represent a possible state of affairs. Since nothing speaks against regarding all these (equivalence classes of) solutions at least as genuine metaphysical possibilities,\footnote{Some people might turn their noses up at those solutions containing closed timelike curves, which they would consider patent metaphysical \emph{impossibilia}. However, in order to be credible, these people should provide some robust argument that justifies their taking general relativity seriously \emph{except for} these solutions. I will gloss over this issue, given that it is orthogonal to the point I want to make in this paper.} our conclusion might have been that \eqref{efe} contains no hint of metaphysically interesting necessity. However, the fine-grained structure of \eqref{efe} tells a slightly different story.

In fact, the geometry/matter coupling is just a part of the laws of general relativity. The rest of the laws take the form \eqref{bian}, but these laws describe a set of \emph{mathematical} relations that the Riemann tensor obeys. Indeed, they were known to mathematicians much before even just Einstein's special theory was published (the standard source is \citealp{570}, but a version of these identities was already derived in \citealp{640}). And here comes the interesting part of the story: given that \eqref{bian} are mathematical identities, they hold by metaphysical necessity, that is, there is no possible world where they are not true; but, if they are integral part of \eqref{efe} in any interesting physical sense (that is, they describe some salient feature of the world), mightn't this mean that they ``drag'' the rest of the laws into a metaphysically necessary status? The next step in my inquiry is to find out whether there is in fact such an interesting physical sense.

\section{The necessary path to Einstein's equations: Misner, Thorne, and Wheeler}\label{sec:3}

A possible reaction to this inciting incident could be to disregard the whole story as evidence that philosophers sometimes tend to overthink things. Agreed, --the objection would go-- \eqref{bian} are part of \eqref{efe}, but so what? Why can't we dismiss \eqref{bian} as a byproduct of the formalism, as we did with the redundant part of \eqref{efe}? After all, the Bianchi identities are just a consequence of a \emph{representational} choice, that is, using the Riemann curvature tensor (which is symmetric) and its covariant derivative. As \citet[][p. 27]{634} puts it, the Bianchi identities are just a ``brute geometrical fact'' about the divergence-freeness of the Einstein tensor (see equation \eqref{consg} below).\footnote{Many thanks to an anonymous reviewer for articulating this objection.}

However, not everybody shares this dismissive attitude. In fact, some of the most prominent physicists who contributed to the development of general relativity since its inception think that the Bianchi identities are an essential part of Einstein's equations in a \emph{physical} sense. Simply speaking, for these people, the Bianchi identities are a key ingredient that makes it possible to couple spacetime to matter. Using John A. Wheeler's metaphor \citep[][pp. 106-107]{589}, the Bianchi identities lie at the root of the ``grip of spacetime'' that ``holds firmly onto the content of [energy-momentum] in every [infinitesimal spacetime region]'', so that it ``bars any creation --or destruction-- of [energy-momentum] anywhere in spacetime''. According to Wheeler, the ``magic'' of this grip shows itself in a principle encoded in the Bianchi identities, i.e. the principle that ``the boundary of a boundary is zero'' (hereafter symbolized by $\partial\partial=0$; see \citealp{27}, chapters 15 and 17, for a thorough presentation, and \citealp{589}, chapter 7, for a less technical but still informative introduction).

\begin{figure}
\begin{center}
\includegraphics[scale=0.35]{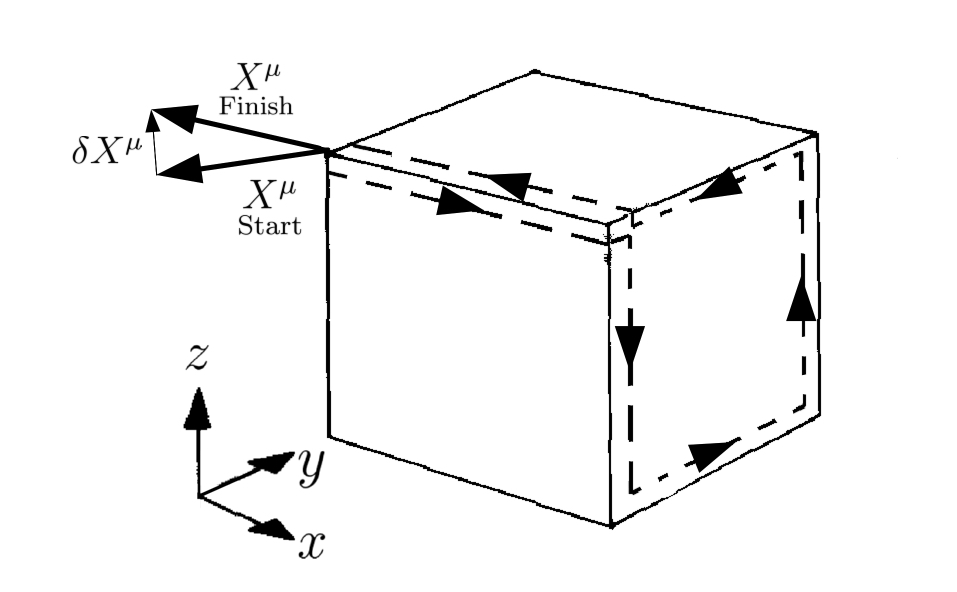}
\caption{Parallel transport of $X^\mu$ around one of $\mathcal{C}$'s faces through its edges.}\label{fig:1}
\end{center}
\end{figure}

To have a rough idea of why the Bianchi identities encode the principle that $\partial\partial=0$, imagine a very tiny (possibly infinitesimal) $3$-cube $\mathcal{C}$ immersed in a generic Riemannian manifold. Consider a vector $X^\mu$ with origin on one edge of $\mathcal{C}$ and parallel transport this vector through the edges of the opposite face of $\mathcal{C}$ back to its initial position (figure \ref{fig:1}).

At the end of the transport procedure, the mismatch $\delta X^\mu$ between the initial and final orientations of $\delta X^\mu$ due to the curvature in that region is (in Riemann normal coordinates\footnote{Cf., for example, \citealp{100}, page 42, for an explication of what these coordinates amount to.}):

\begin{equation}\label{eq:3}
\delta X^\mu=R^\mu_{\phantom{\mu}\nu y z}(\text{at } x+\Delta x)X^\nu\Delta y\Delta z.
\end{equation}

If we instead perform this procedure on the face opposite to the one considered, we find:

\begin{equation}\label{eq:4}
\delta' X^\mu=-R^\mu_{\phantom{\mu}\nu y z}(\text{at } x)X^\nu\Delta y\Delta z,
\end{equation}

where the minus sign stems from the fact that the direction of circulation of $X^\mu$ on this face is opposite to the previous one. By (i) summing \eqref{eq:3} and \eqref{eq:4}, (ii) multiplying both sides by $\frac{\Delta x}{\Delta x}$,  and (iii) recalling how the standard derivative operator is defined,\footnote{$R^\mu_{\phantom{\mu}\nu y z, x}\overset{def}{=}\lim_{\Delta x\to 0}\frac{R^\mu_{\phantom{\mu}\nu y z}(\text{at } x+\Delta x)-R^\mu_{\phantom{\mu}\nu y z}(\text{at } x)}{\Delta x}.$} we find that the total contribution of the two faces is:

\begin{equation}\label{eq:5}
\left(\delta X^\mu\right)_{tot}=R^\mu_{\phantom{\mu}\nu y z, x}X^\nu\Delta x\Delta y\Delta z.
\end{equation}

\begin{figure}
\begin{center}
\includegraphics[scale=0.25]{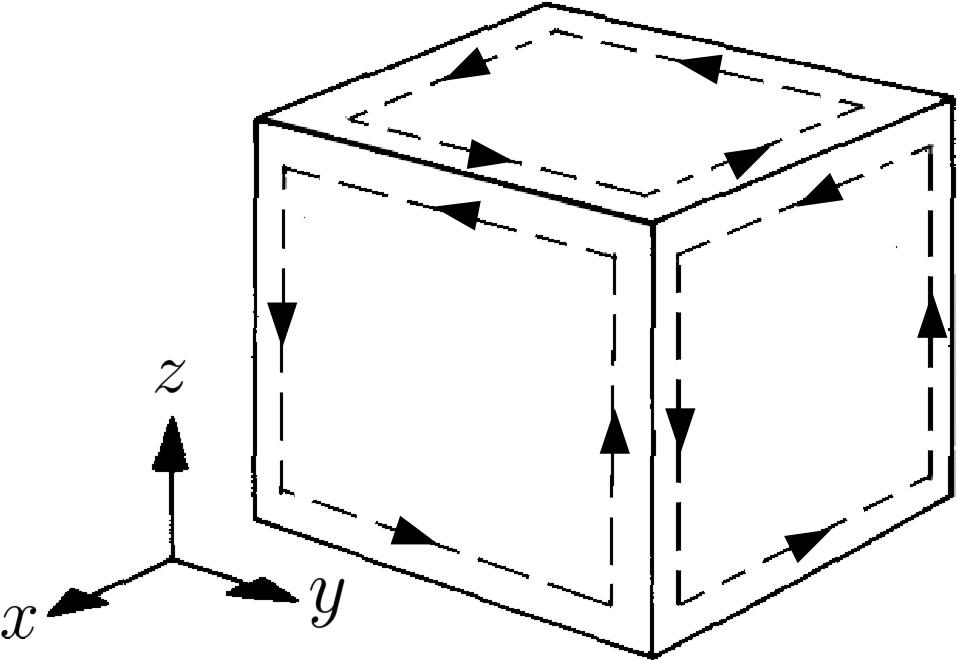}
\caption{Direction of circulation through each face of $\mathcal{C}$. If we choose an arbitrary origin for $X^\mu$ on a corner of $\mathcal{C}$ and we parallel transport it in a circuit through all the six faces of the cube, the vector will pass through each edge twice, once in a direction and once in the opposite one.}\label{fig:2}
\end{center}
\end{figure}

Here comes the key insight. If we perform the parallel transport operation through a circuit that comprises all the six faces of $\mathcal{C}$, we immediately see that $X^\mu$ passes through each edge of the cube twice, once in one direction, once in the opposite one (figure \ref{fig:2}). It is then easy to realize that $\delta X^\mu=0$ in this case, that is, all the curvature-induced changes of direction in $X^\mu$ cancel out. Hence, the sum of the three terms of the form \eqref{eq:5}, corresponding to the contributions of the three couples of opposite faces of $\mathcal{C}$ must add up to zero, which means that:

\begin{equation}\label{eq:6}
R^\mu_{\phantom{\mu}\nu y z, x}+R^\mu_{\phantom{\mu}\nu x y ,z}+R^\mu_{\phantom{\mu}\nu z x ,y}=0.
\end{equation}

If we now want to abandon the Riemann normal coordinates and rewrite \eqref{eq:6} in a generic coordinate system, all we have to do is to substitute ordinary derivatives with covariant ones and swap the $xyz$ indexes with the generic ones $\delta\sigma\rho$. Therefore, in the end, we get exactly the Bianchi identities \eqref{bian}.

At this point, the connection between the Bianchi identities and the principle that $\partial\partial=0$ is very easily established. The elaborate procedure discussed above gives a precise mathematical formulation of the fact that the interior of $\mathcal{C}$ is totally enclosed by the two-dimensional boundary $\partial C$ constituted by its faces, and this is because all the edges of these faces --the one-dimensional boundary of the faces themselves, or $\partial\partial C$-- are pairwise joint together, thus not showing any point to the outside. But the boundary of $\partial C$, by definition, is the set of points whose neighborhoods contain at least a point in $\partial C$ and at least one point outside $\partial C$. Hence, given that the edges have no ``loose'' point (in one dimension), the set $\partial\partial C$ is empty. This result is valid for any closed surface in any dimension, hence the generality of the principle that $\partial\partial=0$.

Now that we grasped the relation between the Bianchi identities \eqref{bian} and the principle that $\partial\partial=0$, we need to clarify why and how such a principle is essential to general relativity. To see this, we start by contracting \eqref{bian} using the metric tensor $g_{\mu\nu}$ to raise and lower the indexes (cf. \citealp{593}, section 3.2, for the actual calculations):

\begin{displaymath}
\begin{array}{l}
g^{\nu \rho}g^{\mu \sigma}\left(R_{\mu  \nu \sigma \rho;\delta}+R_{ \mu \nu \delta \sigma;\rho}+R_{ \mu  \nu \rho \delta;\sigma}\right)=\\
=R_{;\delta}-R^{\rho}_{\phantom{\rho}\delta;\rho}-R^{\sigma}_{\phantom{\sigma}\delta;\sigma}=R_{;\delta}-2R^{\sigma}_{\phantom{\sigma}\delta;\sigma}=0,
\end{array}
\end{displaymath}

from which we get the so-called \emph{contracted} Bianchi identities:

\begin{equation}\label{conbian}
R^{\sigma}_{\phantom{\sigma}\delta;\sigma}-\frac{1}{2}R_{;\delta}=0.
\end{equation}

We can further fiddle with \eqref{conbian} to obtain:

\begin{displaymath}
R^{\sigma}_{\phantom{\sigma}\delta;\sigma}-\frac{1}{2}g^{\sigma}_{\phantom{\sigma}\delta}R_{;\sigma}=\left(R^{\sigma \delta}-\frac{1}{2}g^{\sigma\delta}R\right)_{;\sigma}=0.
\end{displaymath}

It is exactly at this point that general relativity breaks in. Indeed, the expression in parentheses above is just the definition of the Einstein tensor \eqref{eins}, which means that the following holds:

\begin{equation}\label{consg}
G^{\sigma\delta}_{\phantom{\sigma\delta};\sigma}=0.
\end{equation}

In other words, the fact that the principle that $\partial\partial=0$ holds implies that there is a feature of geometry that is conserved (which is some sort of net curvature-induced ``moment of rotation'' associated with a volume element, see \citealp{27}, section 15.3), and such a feature is represented by the Einstein tensor. This result was first proved using the language of exterior calculus by \'Elie Cartan (\citealp{572}, chapter 8, section V 195).

A more pictorial representation of the information encoded in \eqref{consg} and its relation to the principle that $\partial\partial=0$ is as follows (see \citealp{27}, section 15.5, for the technical details). Consider a (infinitesimal) $4$-cube $\mathcal C$ centered around a spacetime point. The boundary of $\mathcal C$, that is, $\partial\mathcal C$, is constituted by eight $3$-cubes (the hyperfaces of $\mathcal C$), each of which has in turn a boundary constituted by six $2$-dimensional faces. Now, in order to calculate how much curvature-induced moment of rotation is created (or destroyed) inside $\mathcal C$, we need to sum the net ``flow'' of such moment through $\partial\mathcal C$, but (because of how this moment of rotation was characterized in the discussion above) this means counting the contribution of each $2$-dimensional face of the eight cubes constituting $\partial\mathcal C$ twice, once with one sign and once with the opposite sign. In the end, the amount of created (or destroyed) moment of rotation inside $\mathcal C$ \emph{has to be} zero by the principle that $\partial\partial=0$, which is what \eqref{consg} states.

By considering \eqref{consg} alongside \eqref{cons}, we finally realize that, if we identify (up to a constant) the moment of rotation inside any infinitesimal volume of spacetime with its energy-momentum content, we automatically get Einstein's equations. Note how such an identification is not pulled out of thin air but is forced upon us by the topological principle that $\partial\partial=0$.\footnote{At least, if we take the formal machinery of the theory as a faithful representation of the world in a specific sense. I will discuss this point in the next section, where I will consider some criticism to this attitude.} This is exactly what Wheeler had in mind when he wrote that the principle that $\partial\partial=0$ is the ``magic'' behind the ``grip'' of spacetime onto the mass-energy content of any infinitesimal spacetime region. It is important to highlight how, in this story, the principle that $\partial\partial=0$ (and the ensuing Bianchi identities) seem to be regarded as concrete facts involving the physical geometry of spacetime, rather than mere mathematical constructions.

But is this moment of rotation/energy-momentum identification choice ``inevitable'', so to speak? In other words, isn't it possible to find other divergence-free rank-2 tensors that depend on the metric tensor in an appropriate way, so that a different identification with the stress-energy tensor can be considered? If that was the case, then much of the ``magic'' Wheeler speaks about would be lost because the principle that $\partial\partial=0$ would not single out \eqref{efe} as the sole choice available.

This issue is settled by the theorem proven in \citet{596, 594}, which shows that, in four dimensions, the only divergence-free rank-$2$ tensor which depends on the metric tensor and its first two derivatives is in fact the Einstein tensor (besides, of course, the metric tensor itself, whose covariant derivative is trivially zero by the requirement of compatibility with the affine connection).\footnote{\citet[][p. 42]{620} asks whether it is in principle possible to obtain an equation of the form \eqref{efe}, where the divergence of the functional on the left does not vanish as a mere matter of identity in $g_{\mu\nu}$. The answer comes from the second Noether's theorem, which implies that this is not possible as long as the gravitational part of the action from which the field equations are derived is generally covariant (see \citealp{621}, section IV, for a discussion of this point).} \citet[][section 7]{677} strengthened this result by proving that it holds for any number of spatiotemporal dimensions if we require that the tensor to be coupled with the stress-energy tensor in fact possess the physical dimensions (not to be confused with the spatiotemporal ones) of stress-energy --meaning that the coupling constant is dimensionless. For clarity's sake, it is important to note that the Bianchi identities are not an explicit premise of Lovelock's theorem, so it is fair to say that Misner, Thorne, and Wheeler's derivation of Einstein's equations relies on the Bianchi identities along with Lovelock's theorem.

There is perhaps no better way to summarize all the above discussion than with the evocative dialogue reported in \citet{27}, p. 364:

\begin{quote}
Physics tells one what to look for: a machinery of coupling between gravitation (spacetime curvature) and source (matter; the stress-energy tensor $\mathbf{T}$) that will guarantee the automatic conservation of the source ($\boldsymbol{\nabla}\cdot\mathbf{T}=0$). Physics therefore asks Mathematics: ``What tensor-like feature of the geometry is automatically conserved?'' Mathematics comes back with the answer: ``The Einstein tensor.'' Physics queries: ``How does this conservation come about?'' Mathematics, in the person of \'Elie Cartan, replies: ``Through the principle that the `boundary of a boundary is zero' ''.
\end{quote}

\section{A necessary connection between spacetime and matter?}\label{sec:new}

At this point, it is quite clear how the strong necessitarians can use all of this to their advantage.  The first step would be to argue that both the Einstein and the stress-energy tensors represent genuine properties borne by spacetime and matter respectively. This step is quite easy to implement (see e.g. \citealp{569}, pages 206-207; but see also \citealp{37}, for a critical discussion of this point), especially considering that $G_{\mu\nu}$ and $T_{\mu\nu}$  are tensorial objects from which we can extract crucial physical information about spacetime and matter (recall the example of the mass-energy density made at the beginning of section \ref{sec:2}). 

The second step would then be to argue that these (sets of) properties are not ``inert'', but bear causal efficacy. A possible example of how this could be done is the following:

\begin{quote}
Each spacetime point is characterized by its dynamical properties, i.e. its disposition to affect the kinetic properties of an object at that point, captured in the gravitational field tensor at that point. The mass of each object is its disposition to change the curvature of spacetime, that is to change the dynamical properties of each spacetime point. Hence all the relevant explanatory properties in this set-up may be characterized dispositionally.\\
(\citealp{384}, p. 240)
\end{quote}

It is important to note that causal theories of properties, including dispositionalism, are often criticized by pointing out that general relativity does not easily accommodate the notion of causation. Here I will gloss over this debate, being content to refer the interested reader to the recent discussion in \citet[][section 2 in particular]{582}, and references therein. Instead, I will focus on an objection to strong necessitarianism --especially dispositional monism-- that has a direct bearing on the debate about laws of nature. This objection is due to Stephen Mumford (see \citealp{242,642}), who is a realist about necessary connections but not about laws --he coined the term \emph{realist lawlessness} to describe his position.\footnote{To be fair, Mumford's original argument targets all kinds of nomological realists. In fact, he argues that nomological realists have to face a ``Central Dilemma'' that leads to conceptual troubles irrespective of which horn is chosen (\citealp{242}, chapter 9). Here I will focus just on the horn that affects strong necessitarians.}

Mumford starts by arguing that laws can be taken metaphysically seriously only insofar as they have an active role in determining the phenomena they are said to govern. For him, claiming that a law $L$ entirely depends on\footnote{Mumford claims that it is immaterial to the argument whether such a dependence is supervenience, reduction, or even constitution (see \citealp{242}, section 9.7).} the entities and events making up the history of a given world $w$ amounts to saying that $L$ does not exist at $w$. This is because $L$ does not determine $w$'s history in any way (it is in fact the other way round), so it can be brushed away from the metaphysical analysis, given that it does not do any relevant explanatory job.

This evidently raises a challenge to strong necessitarian theories that construe laws as dependent on causal properties. The challenge can be summarized by a simple question: how can \emph{real} laws determine the very things they depend on? This challenge particularly impacts dispositionalists \emph{\`a la} Bird, who are nomological realists and claim that laws depend on the dispositional essences possessed by objects. In the present context, Mumford's argument implies that going for a dispositionalist account of spatiotemporal and material properties would undermine the lawhood and necessity of \eqref{efe}.

In order to counter Mumford's challenge, one may try to argue that what laws depend on is in fact distinct from that which is governed. Let us see how this argument may go.\footnote{In the following, I am drawing from Bird's reply to Mumford reported in \citet[][pp. 441-454]{641}, which is a review symposium of \cite{242}.} First of all, it is important to clarify how laws depend on dispositional essences. Simply speaking, if an object $O$ possesses an essentially dispositional property $P$ then, necessarily, it would show a characteristic manifestation $M$ whenever it receives an appropriate stimulus $S$.\footnote{For simplicity's sake, I will not consider chancy dispositions here, given that Einstein's equations are not stochastic. The present discussion is easily generalizable to this type of dispositions though.} This leads to the statement ``for all $O$s, if $O$ has $P$ and receives $S$, then it shows $M$''. This statement has the hallmark of a law-like generalization based on $P$. The dependence of the law on $P$ is quite easy to spot: In order to have a different law, the disposition $P$ must be different as well. Hence, it is plausible to maintain that laws supervene on dispositions.

Now note that (i) dispositions are \emph{distinct} from the set of events in which such dispositions manifest under certain stimuli. Roughly speaking, the existence of dispositions does not depend on the existence of their manifestations, so the manifestation events can be regarded as ``external'' to dispositions. Moreover, (ii) dispositions \emph{determine} the set of manifestation events in a straightforward sense: Trivially, a dispositional essence $P$ \emph{makes it the case} that should $O$ possess $P$ and experience $S$ then there will be $M$.

Since laws supervene on dispositions, (i) implies that manifestation events are external to laws too. Moreover, from (ii), laws acquire a substantial metaphysical import in virtue of the fact that they supervene on something that determines (part of) the world's history --i.e. dispositions governing manifestation events. In this way, the challenge is defused: The dispositional essences of objects guarantee --\emph{pace} Mumford\footnote{Actually, Mumford is not at peace with this response at all. For him, in the above sketched account of laws, all the metaphysically substantial work is done by dispositions, with laws being demoted to some counterfactuals made true by dispositions themselves. Hence, what supervenes on dispositions does not deserve to be called ``law'' (see his reply to Bird reported in  \citealp[][pp. 463-464]{641}).}-- that certain law-like generalizations, such as \eqref{efe}, necessarily hold and do a relevant explanatory job.

If the above response is sound, then the third and final step of the necessitarian strategy would be to show that it is in fact \emph{essential} for spatiotemporal properties to determine the motion of matter and for material properties to change the curvature of spacetime. This step amounts to arguing that the symmetric relation ``$N(G_{\mu\nu}, T_{\mu\nu},)$'' encoded in \eqref{efe}, which conveys the mutual causal behavior of spacetime and matter, holds by metaphysical necessity. Here is where appealing to the whole story told in section \ref{sec:3} would give a huge payoff to the strong necessitarians. According to this story, if we discover by empirical investigation that ``$N(G_{\mu\nu}, T_{\mu\nu},)$'' holds, we have to conclude that it holds by metaphysical necessity. This is because a world where spatiotemporal properties as modelled by $G_{\mu\nu}$ and material properties as modelled by $T_{\mu\nu}$ are not $N$-related is a world where the principle that $\partial\partial=0$ and Lovelock's theorem do not hold, which is an impossible state of affairs. Therefore, if ``$N(G_{\mu\nu}, T_{\mu\nu},)$'' \emph{qua} empirical statement involving the mutual causal behaviour of spacetime and matter is true, this truth is metaphysically necessary. Note how this is a particular case of a Kripkean necessary truth involving natural kinds' characterization: the $N$-relation basically ``defines'' what it is for something to be spacetime or matter.

One might argue that it is conceivable to have counterfactual situations where a slight modification of \eqref{efe} holds, with the value of the coupling constant being a bit different from the actual one. In these situations, $G_{\mu\nu}$ and $T_{\mu\nu}$ would still represent spacetime and matter, but now their causal role would be a bit different from the actual one, which shows that, in fact, the $N$-relation \emph{does not}  define in any interesting metaphysical sense what it is for something to be matter or spacetime (see \citealp{639}, for a general articulation of this anti-necessitarian argument, and \citealp{588}, especially sections 2 and 6, for a reply). Strong necessitarians would react to this argument by pointing out that the coupling constant is an integral part of the relation $N$, which defines the (causal) nature of the spatiotemporal and material properties, so slight modifications of its values --although conceivable-- are not metaphysically possible.

That being said, the opponents of strong necessitarianism would still not find the above sketched strategy particularly compelling. They may reply that the principle that $\partial\partial=0$ and Lovelock's theorem single out general relativity as the only possible metric theory of gravity \emph{under certain specific conditions}. In fact, by relaxing such conditions in one way or another, we end up with a plethora of alternative metric theories of gravity, some of which would count as more or less straightforward \emph{extensions} of general relativity (see \citealp{635}, for a comprehensive review of these approaches, and \citealp{636,637,638}, for a general framework that groups together these extensions as a parametrized family of theories). Why shouldn't these theories be regarded as describing genuine metaphysical possibilities? If we agree with this, then we would have to accept a proliferation of possible worlds where \eqref{efe} strictly speaking do not hold. Some of these worlds would in fact feature sets of properties described by $G_{\mu\nu}$ and $T_{\mu\nu}$, without them being $N$-related. This would be enough at least to ``lower'' the kind of necessitation involved in \eqref{efe} to an Armstrong-like nomic type: In all possible worlds where $G_{\mu\nu}$ represents spatiotemporal properties and $T_{\mu\nu}$ represents material properties \emph{under the appropriate conditions}, the laws of general relativity necessarily hold.

In order to assess the strength of this objection, it is important to clearly state the conditions under which the ``path'' to a metric theory of gravity discussed in the previous section ceases to be the only one practicable. Following the discussion in \citealp{635}, section 2.4.1, we can ``dodge'' the conclusion of Lovelock's theorem, and hence construct a metric theory of gravity different from general relativity, whenever we choose at least one of these options:
\begin{enumerate}
\item\label{pr1} Consider a space with dimensionality different than four.\footnote{Under the light of  the already mentioned result in \cite{677} this is made possible by allowing a dimensionful coupling constant, which would compensate for the fact that the tensor to be equated to the stress-energy one does not have the physical dimensions of stress-energy.}
\item\label{pr2} Consider other degrees of freedom or fields beyond (or rather than) those of general relativity.
\item\label{pr3} Accept higher than second order derivatives of the metric field in the field equations.
\item\label{pr4} Give up on rank-$2$ tensor field equations.
\item\label{pr5} Give up on the symmetry of the field equations under exchange of indices.
\item\label{pr6} Give up on divergence-free field equations.\footnote{A further possibility would be to derive the field equations from an action containing terms that are not functions of the fields or their derivatives evaluated at a single point in spacetime. I will set aside this ``non-locality'' option for simplicity's sake, given that it is not crucial for the discussion.}
\end{enumerate}

We now see that the above objection against strong necessitarians seems quite incisive. While, in fact, strong necessitarians can always respond by saying that each of the options above amounts to denying some essential feature of spacetime or matter\footnote{Perhaps it would be more accurate to claim that these options \emph{misrepresent} spacetime and matter. However, I will gloss over this aspect, given that it is not central to the point at stake. The reader interested in the topic of scientific representation can refer to \citet{647}, and references therein.} (thus constituting a metaphysically impossible condition), it is evident that this requires from them a lot more argumentative effort than that implied \emph{prima facie} by the three-step strategy sketched at the beginning of this section. In other words, the above objection shows that strong necessitarians have to accept and justify quite a number of carefully tailored essential features of reality in order to make sure that the options (\ref{pr1}) to (\ref{pr6}) are not metaphysically viable.

Even if the above criticism sounds penetrating, still nothing prevents the strong necessitarians from shrugging it off just by biting the bullet. Agreed, their reply might go, in order to avoid (\ref{pr1}) to (\ref{pr6}) from being genuine possibilities one must argue that many facts regarding spacetime and matter (e.g. spacetime having four dimensions) are necessary \emph{a posteriori}. But so what? The physical world is extremely complex, and it encompasses an exorbitant amount of facts. So, even if just a small subset of these physical facts are necessary, still they will very likely be a huge number. In particular, the fact that spacetime and matter possess many essential features --which bar (\ref{pr1}) to (\ref{pr6}) from being metaphysically viable options--  is no surprise given the tremendous amount of structure that is grouped under the terms ``spacetime'' and ``matter'', so it would be unfair to accuse the strong necessitarians of inflating and fine-tuning their metaphysics in order to render modified gravity theories metaphysically impossible. Therefore, the most that the above objection shows is that the strong necessitarians have some work to do in order to defend their view, not that such a view is inherently untenable or \emph{ad hoc}.

In order to get a better grasp of the kind of argumentative work that the strong necessitarians have to go through, let us focus on a concrete example. Consider option (\ref{pr6}) above: This option is compatible with claiming that there are possible worlds where the conservation law \eqref{cons} for the field source does not hold and, hence, the discussed coupling of spacetime and matter made at the level of $G^{\mu\nu}_{\phantom{{\mu\nu}};\mu}=T^{\mu\nu}_{\phantom{{\mu\nu}};\mu}=0$ cannot be established there. 

The strong necessitarians can challenge this claim by mentioning a result due to Dirac (see \citealp[][section 30]{622}; \citealp[][section V]{621}), which in a nutshell amounts to saying that the condition \eqref{cons} is automatically fulfilled by any type of material source once we require that the material part of the action\footnote{The other part being the gravitational one.} -- from which we get the field equations/equations of motion for our theory by extremization -- has to be generally covariant. Most importantly, this result does not depend on the form of the action itself, but just on its covariance under arbitrary coordinate transformations. 

It is easy to see what it is, which the Dirac's result highlights, and which the strong necessitarians can use to strengthen their case. Simply speaking, the condition \eqref{cons} is so weak that coming up with a state of affairs that violates it would involve an extremely malicious and \emph{ad hoc} tweaking of a world.\footnote{As \citet[][section 5-1]{599} clearly points out, the condition $T^{\mu\nu}_{\phantom{{\mu\nu}};\mu}=0$ holds even when stronger conservation laws for $T^{\mu\nu}$ --either differential or integral-- do not. Hence, it may very well be the case that \eqref{cons} still holds in a ``crazy'' world where nothing is in fact conserved in the stronger sense usually adopted in physics.} For the strong necessitarian, this would be a red flag signalling that we are messing up with something essential about matter (and spacetime).

Against this strong necessitarian line of reasoning based on Dirac's result, one might question the physical import of the requirement of general covariance for the action. This position goes back to \citet{59}, who showed that any theory can be rendered generally covariant with some appropriate mathematical manipulations. To this, the strong necessitarians can react first of all by questioning whether general covariance in general relativity is really devoid of physical import. The status of general covariance in general relativity is in fact a \emph{vexata quaestio} in the philosophy of spacetime physics  (see, e.g. \citealp{365,370}, to catch a glimpse of the debate; but see also \citealp{373} for an attempt at reconciling the opposing parties), but the important point here is that strong necessitarians can endorse the claim that general covariance in general relativity is physically substantive in the sense that it expresses a local gauge freedom of the theory -- which makes general relativity a full-fledged gauge theory of gravity.

By going for the ``general relativity as a gauge theory of gravity'' story, the strong necessitarians would add a further arrow to their quiver against those who question the physical import of the Bianchi identities. In a nutshell, gauge theories are field theories (classical and quantum) whose dynamics is encoded in a Lagrangian (from which the action is constructed) which is invariant under a group of local (that is, depending on a number of arbitrary functions of spacetime) transformations specific for each theory. The structure of gauge theories has proved itself to be extremely powerful, to the point that the entirety of modern fundamental physics successfully describes the world by means of gauge theories (see \citealp{148} for a nice philosophical discussion, and \citealp{623}, for a brief but enlightening historical treatment). Now, as Noether's second theorem shows, any theory which is invariant under a continuous group of transformations depending on $n$ arbitrary functions of spacetime exhibits an interdependence of its field equations/equations of motion encoded in $n$ differential identities, which implies that any solution of said equations are determined up to $n$ freely specifiable functions (cf., for example, \citealp[][section 5-2]{599} and \citealp[][section 5.2]{624}). The particular case which involves the general covariance group with $n=4$ is nothing but the case of general relativity, the $4$ differential identities being exactly the Bianchi identities \eqref{bian}. Noether's result together with the huge empirical success of gauge theories are, in the eyes of the strong necessitarians, a further clue of the physical significance and metaphysical inevitability of the story told in section \ref{sec:3}. Also in this case, the opponents might claim that it is just a contingent fact that gauge theories are successful in our world, while the strong necessitarians would defend the thesis that a possible world has to possess a gauge-theoretic structure by using arguments similar to those used to argue that, say, water has to possess its actual molecular structure. From this point on, the challenge returns to a purely metaphysical battlefield, which we are not interested to step on in this paper.

Instead, another metaphysically interesting yet physically-related aspect of the debate that still needs to be addressed is the following: Is it possible to construe general relativity in a way that sidesteps the ``necessary path'' walked by Misner, Thorne, and Wheeler (and Cartan, before them)?

\section{Einstein's equations from a contingent state of affairs: Ehlers, Pirani, and Schild}\label{sec:4}

Many readers may object that the presentation in section \ref{sec:3} looks awkwardly upside-down. For them, that section purportedly shown how Einstein's equations are immediately established as a matter of geometry once the weak conservation of the source of spacetime curvature is assumed, whereas in fact it is the exact opposite: If the laws of general relativity hold, then the coupling of matter with spacetime can be given the nice geometric representation elicited by Cartan's work. This skeptic attitude naturally fits a Humean reading of the laws of general relativity.\footnote{As in the case of necessitarianism, also Humeanism comes in different flavors. Among these different approaches, it is worth mentioning \citet{616,643,485}. Here I will brush over these differences, given that they are not vital for the discussion.} According to this reading, \eqref{efe} supervene on the mosaic of matters of particular fact that contingently obtain at our world, in the sense that these laws belong to the simplest yet most informative deductive system that successfully describes  the mosaic (the so-called \emph{best system}). This point of view is radically anti-necessitarian in that it does not presuppose any modal connection whatsoever among the facts in the mosaic: Everything might have been otherwise. Under this picture, the hint of necessity encoded in the Bianchi identities is totally washed away, in the sense that \eqref{bian} \emph{qua} part of the physical law \eqref{efe} hold just in virtue of it being a theorem of the best system for a general relativistic world. In other words, \eqref{bian} do not ``force'' things to be in a certain way; on the contrary, the fact that some contingent state of affairs obtains makes it possible for them to be best described in terms of \eqref{bian}. In the physical literature there are several approaches that seek to construe general relativity --and especially its geometric machinery-- from some underlying non-inherently geometric states of affairs (for example, \citealp{645}, and \citealp[][chapter 16]{644}, attempt at deriving general relativity from thermodynamic phenomena). Here I will focus my attention on a particular framework put forward by the physicists Ehlers, Pirani, and Schild \citep{577}, which has the remarkable feature of being cast in terms of a deductive axiomatic system, thus making its Humean reading rather straightforward.

In a nutshell, the authors propose an axiomatic system based on the primitive notions of light ray and freely falling particle, which is able to recover the Riemannian geometry of general relativity. It is important to note that this choice of primitives is not conceptually forced upon us by the need to recover Riemannian geometry: For example, \citet[][especially chapters II and III]{601} suggests to take the notions of particle and clock as primitives. Going back to Ehlers, Pirani, and Schild, their approach supplies a list of axioms, in their words, ``suggested by experience'' \citep[][section 2]{577} that the set of light rays $L$ and the set of freely falling particles $P$ (being two subsets of the set of all events $M$) have to obey in order to define, respectively, a conformal and a projective structure over $M$. The first structure permits to define the notions of timelike, lightlike, and spacelike vectors (infinitesimal light cone structure), while the second supplies a notion of parallel transport and, hence, of affine geodesic. Another axiom requires these two structures to be compatible, that is, that all light rays are (lightlike) geodesics. The set $M=(L,P)$ endowed with these two compatible structures is called by the authors \emph{Weyl space} (thus acknowledging the seminal work of the German mathematical physicist; see, e.g., \citealp{602}). The rest of their work is meant to show how, by supplying some more axiomatic conditions, a Weyl space can be reduced to a Riemannian space with a full metric structure.

Here I am not concerned with the technical details of Ehlers, Pirani, and Schild's framework. Rather, I am interested in establishing a connection between their work and the Humean framework. Such a connection is indeed easy to establish: Their work suggests how to get the machinery of Riemannian geometry out of a mosaic of material particles' and photons' trajectories. This is exactly what the Humeans were searching for: A way to show that Riemannian geometry is not inherent into the physical world --thus (modally) constraining facts within it-- but, instead, it is a useful tool to describe the contingent happenings in the mosaic. Thus, for example, there can be ``crazy'' possible worlds where this description of the mosaic in geometric terms is not viable. This is enough for the Humeans to resist the necessitarian push that comes with Wheeler, Thorne, and Misner's story about general relativity. 

Just to be clear, this is not to claim that the Humeans want to eschew spatiotemporal properties and relations from the mosaic -- to the contrary, they firmly believe in the inherent spatiotemporality of worlds like ours (to appreciate the pivotal role that spatiotemporal relations play in the Lewisian/Humean framework, see in particular \citealp[][section 1.6]{284}). Instead, what the Humeans want to eradicate from the picture is any hint of geometric necessity associated to the spatiotemporal nature of the mosaic. Hence, say, they are totally willing to accept that three material objects $A,B,C$ inhabiting our world are spatially related so that their distances fulfill the triangle inequality; what they resist is instead the claim that it might have not been the case that $A,B,C$ (co-)existed yet they were not related in a way satisfying the triangle inequality. In fact, they would claim, far away from our modal horizon there is a rather strange world where this is exactly the case. Such a world might not even feature spatiotemporal relations properly said, as long as they are substituted by ``spatiotemporally analogical'' relations which are --as discussed in \citet[][pp. 75-76]{284}-- (i) natural (i.e. not gerrymandered), (ii) pervasive (in the above example, if $A$ is related to $B$, and $B$ to $C$, then there is also a relation linking $A$ and $C$ directly), (iii) discriminating (if a possible world is large enough, then the relations may be enough to individuate uniquely the \emph{relata}), and (iv) external (i.e. they do not supervene on the intrinsic features of the \emph{relata} taken individually). Note how none of these four minimal requirements presupposes or implies the triangle inequality.

That being said, a moment of reflection shows that, in the context of general relativity, the Humeans might have an Achilles' heel.\footnote{To my knowledge, this worry was firstly voiced in \citet[][section 5]{469}.} Such a potential vulnerability is indirectly highlighted by the assessment that Ehlers himself gave of his framework:

\begin{quote}
This approach shows how quantitative measures of time, angle and distance, and a procedure of parallel displacement [...] can be obtained constructively from `geometry-free' assumptions about light-rays and freely falling particles; pseudo-Riemannian (or Weylian) geometry is recognized even more clearly than before as the appropriate language \emph{for a generalized kinematics} which allows for the unavoidable and ever-present `distortions' called gravitational fields.\\
\citep[][p. 81, my emphasis]{604}
\end{quote}

Otherwise said, the above sketched procedure yields single models $\langle g_{\mu\nu}, T_{\mu\nu}\rangle$ of \eqref{efe}, depending on the particular arrangement of trajectories, but it is not clear whether it captures the so-called \emph{background independence} of the theory --i.e. the fact that spacetime is dynamical, and not just a fixed arena where the dynamics of matter unfolds (see, e.g., \citealp{47}, for a in-depth analysis of this tricky concept).

To have a better idea of the issue at stake, we can formulate it as a simple question: Is it possible for the Humeans to capture the background independence of general relativity by looking at the (entire) mosaic? In other words, granted that the Humeans can recover the specific geometry $\tilde{g}_{\mu\nu}$ of spacetime and the specific distribution $\widetilde{T}_{\mu\nu}$ of matter from the mosaic obtaining at a world $w$, are they able to discern whether $w$ is a cosmological model $\langle \tilde{g}_{\mu\nu}, \widetilde{T}_{\mu\nu}\rangle$ of general relativity (where spacetime is a ``dynamic partner'' of matter) or a world in which the material distribution $\widetilde{T}_{\mu\nu}$ just inhabits a fixed background that happens to be $\tilde{g}_{\mu\nu}$? If the Humeans cannot answer these questions in the positive, this might hint at the fact that, after all, it is not the full laws of general relativity that supervene on a mosaic but, at most, just a particular instance of them. It is easy to see that such a potential problem for the Humeans does not stem from Ehlers, Pirani, and Schild's approach to general relativity \emph{per se} but it is built into the theory, so to speak. The necessitarians would be happy to point out that such a problem does not arise if we introduce genuine modal features --e.g. in the guise of causal properties-- in the mosaic. In this way, in fact, the co-variation of spacetime and matter encoded in \eqref{efe} is easily accounted for.

The Humeans can defuse this challenge by pointing out that, if $w$ is a general relativistic world, then $\tilde{g}_{\mu\nu}$ and $\widetilde{T}_{\mu\nu}$ cannot be the simplest and most informative descriptions of matters of fact in the mosaic at $w$ without the correlations underlying \eqref{efe} being part of the best system as well. Let us consider this response in more detail.

In physically realistic situations, we expect $\tilde{g}_{\mu\nu}$ and $\widetilde{T}_{\mu\nu}$ to have very complicated forms (imagine how complex the detailed description of the geometry of our world might be), so these mathematical formulae are viable only insofar as they are the simplest and strongest descriptions that make it possible for the best system at $w$ to express the entailment of counterfactuals of the form ``had the particles' and photons' trajectories been distributed in such and such a way, the spacetime geometry and energy-momentum distribution would have been such and such''. Note that it is these counterfactuals that supply the kind of information that captures the co-variation of spacetime and matter encoded in \eqref{efe}. Now, if the best system at $w$ did not entail any such counterfactuals (or just some of them), thus failing to capture \eqref{efe}, then the simplicity and strength of $\tilde{g}_{\mu\nu}$ and $\widetilde{T}_{\mu\nu}$ would become dubious: At that point, for example, why shouldn't a much simpler metric and a more complicated stress-energy tensor be the simplest and strongest choices overall? Of course, this does not rule out the possibility of a world $w'$ where the uncorrelated $\tilde{g}_{\mu\nu}$ and $\widetilde{T}_{\mu\nu}$ are in fact part of the best system; but such a world would be much ``messier'' than $w$ (and any other general relativistic world) in an empirically detectable way. In short, the fact about whether $\tilde{g}_{\mu\nu}$ and $\widetilde{T}_{\mu\nu}$ are correlated or not has to boil down to empirical facts about the mosaic that the best system is certainly able to capture in the form of the above mentioned counterfactuals, so it is out of question that there can be an ambiguity between $\langle \tilde{g}_{\mu\nu}, \widetilde{T}_{\mu\nu}\rangle$-correlated and $\langle \tilde{g}_{\mu\nu}, \widetilde{T}_{\mu\nu}\rangle$-uncorrelated at a given world (at least, in physically realistic situations).

Obviously, the fact that the best system at $w$ entails certain counterfactuals does not mean that some primitive modal notion is being smuggled into the mosaic at $w$. The Humeans have no problem in grounding counterfactual reasoning in inherently non-modal facts (e.g., in an ontology of individuals endowed with non-modal properties), and this can be equally achieved by modal realists \emph{\`a la} Lewis \citep[][section 1.2]{284} as well as modal fictionalists \emph{\`a la} Divers \citep{646}. It has to be pointed out, however, that accounting for counterfactual reasoning in general relativity may be tricky, since the absence of fixed background spatiotemporal structures that ``persist'' across possible worlds\footnote{For example, all possible worlds in which special relativity holds feature Minkowski spacetime, so this structure ``persists'' across this cluster of worlds. See, e.g., \citet[][section 3]{467} for a characterization of background structures in terms of possible worlds.} makes it difficult to establish a reliable reference for trans-world identification of things (be it material objects or spatiotemporal points and regions), which is required to assess counterfactual change (see \citealp{487}, for a clear articulation of the problem, and \citealp[][sections 4 and 5]{582}, for an alternative framework for counterfactual reasoning that may be viable in general relativity). Clearly, this particular issue impacts also the necessitarians, whose characterization of causal properties involves counterfactual reasoning.

In conclusion, there seems to be no \emph{prima facie} reason to think that the Humeans may be in trouble with the background independence of general relativity. Hence, the onus is on the necessitarians to show that there can be general relativistic mosaics that fail to pick out \eqref{efe} unless some primitive modal features are introduced in it.

\section{Conclusion}
General relativity surely represents a favorable environment for necessitarians about laws of nature. In particular, strong necessitarians can very much profit in defending their views from the ``metaphysical rigidity'' that the theory brings into the world by virtue of geometrical facts becoming physical facts in the strong sense entailed by \eqref{efe}. The most patent example is that of the Bianchi identities, which general relativity seems to promote from mathematical to physical truths entailing the conservation of a physical feature of reality described by the Einstein tensor. However, as we have seen in section \ref{sec:new}, implementing a strong necessitarian strategy that exploits Misner, Thorne, and Wheeler's ``necessary path'' to general relativity is not as straightforward as one might have expected.

Of course, these necessitarian efforts do not move those metaphysicians more rooted in the empiricist tradition. For them, the necessitarians' enthusiasm just stems from taking too seriously --in a quasi Platonist fashion-- the formal machinery of general relativity and gauge theories in general, which heavily relies on differential geometry. The empiricist skepticism is rather simple: Since all measurements always boil down to observations of material facts, and never of purely geometric facts, there must be a way to show that geometry is just a useful way to describe the behavior of material systems. The pulp of this skepticism is usually enclosed in a Humean shell. As discussed in section \ref{sec:4}, the Humeans can indeed point out that the laws of general relativity can be derived from contingent states of affairs with no hint of geometric necessity in them. Necessitarians may try to undermine the Humeans' confidence by devising some malicious cases in which \eqref{efe} fail to supervene on such a mosaic but, as things stand, it is not clear if and to what extent these cases may be really problematic.

In the end, even if the discussion carried out in this paper does not decisively shift the metaphysical balance towards any of the parties involved, still it highlights how reflecting on the nature of Einstein's equations helps sharpening and deepening the broader debate about the laws of nature.

\pdfbookmark[1]{Acknowledgements}{acknowledgements}
\begin{center}
\textbf{Acknowledgements}:
\end{center}
I am indebted to Carl Hoefer for some enlightening discussions about laws of nature in general relativity. I am also grateful to two anonymous reviewers for their comments on earlier versions of the manuscript. I worked on the very first draft of this paper while being at the University of Barcelona as a Juan de la Cierva Fellow, hence I acknowledge financial support from the Spanish Ministry of Science, Innovation and Universities, fellowship IJCI-2015-23321. I carried out the remaining part of the writing process at the Warsaw University of Technology with financial support from the Polish National Science Centre, grant No. 2019/33/B/HS1/01772.

\pdfbookmark[1]{References}{references}
\bibliography{biblio}

\end{document}